\documentclass[12pt]{article}
\def\bd{\begin{displaymath}}\def\ed{\end{displaymath}}
\def\be{\begin{equation}}\def\ee{\end{equation}}
\def\bea{\begin{eqnarray}}\def\eea{\end{eqnarray}}
\def\ba{\begin{array}}\def\ea{\end{array}}
\def\lb{\label}

\def\b{\beta}\def\d{\delta}\def\e{\epsilon}
\def\f{\phi}\def\h{\theta}
\def\o{\omega}\def
\p{\pi}\def\q{\psi}\def\r{\rho}
\def\x{\xi}

\def\O{\Omega}

\def\de{\partial}
\def\inf{\infty}\def\id{\equiv}\def\mo{{-1}}

\def\ex{{\rm e}}

\def\coo{coordinates }

\def\hr{Hawking radiation }

\def\rep{representation }

\def\pb{Poisson brackets }

\def\cor{commutation relations }

\def\PRL#1{Phys.\ Rev.\ Lett.\ {\bf#1}}
\def\PR#1{Phys.\ Rev.\ {\bf#1}}\def\CQG#1{Class.\ Quantum Grav.\ {\bf#1}}
\def\NP#1{Nucl.\ Phys.\ {\bf#1}}
\def\JMP#1{J.\ Math.\ Phys.\ {\bf#1}}\def\PTP#1{Prog.\ Theor.\ Phys.\ {\bf#1}}

\def\JoP#1{J.\ Phys.\ {\bf#1}} 
 
\def\PRep#1{Phys.\ Rep.\ {\bf#1}}

\def\JHEP#1{JHEP\ {\bf#1}}
\def\RMP#1{Rev.\ Mod.\ Phys.\ {\bf#1}}

\def\cD{{\cal D}}\def\ubp{1+\b^2p^2}\def\ver{\,|\,}
\def\hH{{\hat H}}\def\hp{{\hat p}}\def\hq{{\hat q}}\def\hr{{\hat\r}}\def\hJ{{\hat J}}

\begin{document}
\begin{titlepage}
\title{Path integrals in Snyder space}
\vskip60pt
\author{ S. Mignemi and R. \v Strajn\medskip\\ \small Dipartimento di Matematica e Informatica,\\
\small Universit\`a di Cagliari, viale Merello 92, 09123 Cagliari, Italy\\
\small and INFN, Sezione di Cagliari, Cittadella Universitaria, 09042 Monserrato, Italy}
\date{}
\maketitle
\vskip60pt
\begin{abstract}
The definition of path integrals in one- and two-dimensional Snyder space is discussed in detail
both in the traditional setting and in the first-order formalism of Faddeev and Jackiw.

\end{abstract}

\end{titlepage}

\section{Introduction}
The interest in noncommutative spaces has increased in recent years, because they
may describe the structure of space (or spacetime) at the Planck scale, as
several approaches to quantum gravity seem to indicate \cite{nc}.
The formulation of quantum mechanics on a noncommutative space is usually called
noncommutative quantum mechanics. Path integral techniques have demonstrated to be convenient
in the study of this theory.

A characteristic of noncommutative spaces is that the corresponding classical phase space is not
canonical, i.e.\ the \pb do not have the usual form.
However, the standard definition of path integral assumes a canonical phase space \cite{FH,kl},
and one has therefore to extend the formalism to include this more general situation.

This is an interesting problem, that has been afforded in a variety of ways \cite{ac,ncpi}. However,
most work on the subject has been developed for the so-called Moyal plane \cite{mo}, a simple model
whose \pb are constant tensors, hence necessarily implying the breakdown of the Lorentz invariance.

However, more general models of noncommutative spaces exist, in which the Lorentz invariance is
preserved. The best known is the Snyder model \cite{sn}, which, in spite of the presence in its
definition of a parameter $\b$ with the dimension of inverse momentum, is Lorentz invariant.
The quantum mechanics
of the Snyder model has been studied in several papers \cite{mi,LS}.

In its nonrelativistic version, the Snyder model is based on a deformation of the Heisenberg algebra,
given by the \cor
\be
[q_i,p_j]=i(\d_{ij}+\b^2p_ip_j),\qquad[q_i,q_j]=i\b^2J_{ij},\qquad[p_i,p_j]=0,
\ee
where $q_i$ and $p_i$ are the
phase space coordinates, and $J_{ij}$ are the angular momentum
generators; we use units in which $\hbar=1$. Clearly, the classical limit of these commutators
gives rise to a noncanonical phase space.

In this paper, we discuss the formulation of the nonrelativistic quantum mechanics
of the Snyder model through path integral methods, following the approach of \cite{ac}.
We also obtain the same results using the more formal techniques introduced in \cite{FJ} for
the study of first-order systems.

Recently, this problem has also been investigated in \cite{val} in the case of one spatial dimension.

\vfill\eject

\section{Noncanonical classical mechanics}
Before discussing the path integral formulation of Snyder quantum mechanics, we shortly review some
facts concerning noncanonical Hamiltonian formalism \cite{san,jac}, that will be useful in the following.

Let us consider noncanonical fundamental \pb
\be
\{\x_i,\x_j\}=\O_{ij}(\x),
\ee
where $\x_i$ denotes the phase space variables  $q_i$ and $p_i$ and $\O_{ij}$ is an invertible matrix.
Then the Hamilton equations for the Hamiltonian $H(\x)$ read
\be\lb{haeq}
\dot\x_i=\O_{ij}{\de H\over\de\x_j},
\ee
or equivalently,
\be
(\O^\mo)^{ij}\dot\x_j={\de H\over\de\x_i},
\ee

We want to obtain these equation from the variation of a first-order action of the form
\be\lb{fac}
I=\int[a^i(\x)\dot\x_i-H(\x)]dt.
\ee
Then the condition
\be\lb{sympl}
{\de a^j\over\de\x_i}-{\de a^i\over\de\x_j}=(\O^\mo)^{ij}
\ee
must hold. Solving (\ref{sympl}) for the $a^i$, one can write down the action which generates the Hamilton
equations (\ref{haeq}).

\section{One-dimensional Snyder path integral}
Of course, when investigating the Snyder model, one must use the phase space formulation of the path integral.
For a particle satisfying canonical Poisson brackets, moving in a one-dimensional space, this is given by
\be\lb{pi}
A=\int\cD p\cD q\ \ex^{iI},
\ee
where
\be
I=\int_{t_i}^{t_f}Ldt=\int_{t_i}^{t_f}(p\dot q-H(q,p))dt
\ee
is the action (with $L$ the Lagrangian and $H$ the Hamiltonian), and $\cD p\cD q$ is
a measure on the space of paths on phase space that will be defined below.

It can be shown that in a momentum basis the transition amplitude from an initial state
of momentum $p_i$ at time $t_i$ to a final state of momentum $p_f$ at time $t_f$ is given by
\be
<p_f|e^{-i\hH(t_f-t_i)}|p_i>\ =A.
\ee
We have chosen a momentum basis, because, when we shall consider Snyder space, the standard position variables
will not commute and hence do not form a complete set of observables.

We wish to generalize this formula to the one-dimensional Snyder phase space (see also \cite{val}),
whose only nontrivial Poisson bracket is
\be\lb{1dpb}
\{q,p\}=1+\b^2p^2.
\ee
Given the Hamiltonian $H={p^2\over2}+V(q)$, the Hamilton equations in Snyder space read
\be
\dot q=(\ubp)p,\qquad\dot p=-(\ubp){\de V\over\de q}.
\ee

These equations can be obtained from an action principle, by modifying the definition of the
Lagrangian so that (\ref{1dpb}) holds. Using the procedure of section 2, one gets the action
\be\lb{1dac}
I=\int\left(-{q\dot p\over\ubp}-H\right)dt=\int\left({\arctan\b p\over\b}\,\dot q-H\right)dt,
\ee
where the two expressions are equivalent modulo an integration by parts.
We will now show that inserting (\ref{1dac}) into (\ref{pi}) gives the correct expression for
the path integral.

We first recall some results concerning the quantum mechanics of the one-dimensional Snyder model
\cite{KM}.
The Poisson bracket (\ref{1dpb}) goes into the commutator
\be\lb{1dco}
[\hq,\hp]=i(1+\b^2\hp^2).
\ee
The operators $\hq$ and $\hp$ obeying (\ref{1dco}) can be represented in a momentum basis by \cite{HT}
\be
\hp=p,\qquad\hq=i(1+\b^2p^2){\de\over\de p}.
\ee
These operators are hermitian with respect to the scalar product
\be
<\q\ver\f>=\int_{-\inf}^{+\inf}{dp\over\ubp}\ \q^*(p)\f(p).
\ee
The identity operator can therefore be expanded in terms of momentum eigenstates $\ver p>$ as
\cite{KM}
\be\lb{comp}
1=\int_{-\inf}^{\inf} {dp\over\ubp}\ver p><p\ver,\quad{\rm with}\quad<p\ver p'>=(1+\b^2p^2)\d(p-p').
\ee

The eigenvalue equation for the position operator, $\hq\ver q>=q\ver q>$, has formal solutions\footnote{
These eigenstates are not physical, because they have infinite energy \cite{KM}, but are sufficiently regular
to adopt them in this setting.}
\be\lb{1dpq}
<p\ver q>\ \propto\,\ex^{-iq\,{\arctan\b p\over\b}}
\ee
These states form an overcomplete set. However, one can choose a discrete basis with $q=2\b n$, $n$ integer,
which satisfies the completeness relation \cite{KM}
\be\lb{comq}
1=\int dq\ver q><q\ver,\quad{\rm with}\quad<q\ver q'>=\d_{qq'}.
\ee
For simplicity of notation, we have used an integral sign for the infinite sum over $n$.

Let us go back to the path integral.
Splitting the interval $t_f-t_i$ into $N$ intervals of length $\e=t_k-t_{k-1}$,
the standard definition reduces in our case to
\be
A=\int\prod_{k=1}^{N-1}{dp^{(k)}\over1+\b^2p^{(k)2}}\int\prod_{k=1}^Ndq^{(k)}\prod_{k=1}^{N}<p^{(k)}\ver q^{(k)}>\,
<q^{(k)}|e^{-i\e\hH}|p^{(k-1)}>,
\ee
where the completeness relations (\ref{comp}) and (\ref{comq}) have been used. Recalling (\ref{1dpq}), one obtains
\be
<q\ver e^{-i\e\hH}\ver p>\ \propto\,\ex^{iq\,{\arctan\b p\over\b}-i\e H}
\ee
Finally, in the limit $\e\to0$, taking into account that
$p_k-p_{k-1}\sim\dot p\,\e\to\dot p\,dt$, one recovers for $A$ the form (\ref{pi})
with
\be
I=\int\left[-{q\dot p\over\ubp}-H(q,p)\right]dt
\ee
and
\be\lb{mes1}
\cD p=\lim_{N\to\inf}\prod_{k=1}^{N-1}{dp^{(k)}\over1+\b^2p^{(k)\,2}},\qquad\cD q=\lim_{N\to\inf}\prod_{k=1}^{N}dq^{(k)},
\ee
which proves our claim.
\section{Two dimensional Snyder path integral}
In higher dimensions the problem is more difficult, since the position operators $\hq_i$ do not commute and their
eigenfunctions cannot be taken as
a basis for the Hilbert space. It is convenient to use radial \coo instead. In particular, we shall discuss the
case $D=2$, for which
\be
\{q_i,p_j\}=\d_{ij}+\b^2p_ip_j,\qquad\{q_i,q_j\}=\b^2J_{ij},\qquad\{p_i,p_j\}=0,
\ee
where $J_{ij}=p_jq_i-p_iq_j$.
We choose polar \coo to parametrize the momentum space,
and their canonically conjugate variables for the position space. More precisely, we define
\be
p_\r=\sqrt{p_1^2+p_2^2},\qquad p_\h=\arctan{p_2\over p_1},
\ee
and
\be
\r={p_1q_1+p_2q_2\over\sqrt{p_1^2+p_2^2}},\qquad J=J_{12}=p_2q_1-p_1q_2.
\ee
The Jacobian of the transformation is 1. Note that classically $r^2\id q_1^2+q_2^2=\r^2+J^2/p_\r^2$.

The position \coo $\r$ and $J$ essentially correspond to the parallel and orthogonal components of the position vector
with respect to the momentum of the particle.
The phase space polar \coo defined above obey the \pb
\bea\lb{ppb}
&&\{p_\r,p_\h\}=0,\qquad\quad\ \{\r,J\}=0,\qquad\quad\ \{p_\r,J\}=0,\cr
&&\{p_\r,\r\}=1+\b^2p_\r^2,\qquad\{p_\h,J\}=1,\qquad\{p_\h,\r\}=0.
\eea
Notice that the Poisson bracket of $\r$ and $J$ vanishes, and hence the corresponding quantum operators commute and
form a basis for the position space. An alternative basis would be constituted by the operators $\hat r^2$ and
$\hat J$, where $\hat r^2\id\hq_1^2+\hq_2^2$ \cite{LS}.  These \coo however give rise to more complicated formulas.

Again, by the methods of section 2, one can obtain the action that generates the classical Hamilton equations
through the \pb (\ref{ppb}).
This is
\be\lb{2dac}
I=-\int\left[{\r\dot p_\r\over1+\b^2p_\r^2}+J\dot p_\h+H\right]dt=\int\left[{\arctan\b p_\r\over\b}\,\dot\r
+p_\h\dot J-H\right]dt
\ee
where
\be
H={p_\r^2\over2}+V(\r,J).
\ee

Let us now consider the quantum theory. The quantum operators must be defined carefully, because of ordering ambiguities.
We adopt an ordering with $\hp_i$ always on the left. The \cor are
\be\lb{tdcr}
[\hq_i,\hp_j]=i(\d_{ij}+\b^2\hp_i\hp_j),\qquad[\hq_i,\hq_j]=i\hJ_{ij},\qquad[\hp_i,\hp_j]=0
\ee
A \rep of (\ref{tdcr}) is given by \cite{HT}
\be
\hp_i=p_i,\qquad\hq_i=i\left({\de\over\de p_i}+\b^2p_ip_j{\de\over\de p_j}\right)
\ee

The hermitian operators corresponding to the classical polar \coo are
\be
\hp_\r=\sqrt{p_i^2}\id p_\r,\qquad\hp_\h=\arctan{p_2\over p_1}\id p_\h,
\ee
and
\be
\hr=i(1+\b^2p_\r^2)\left({\de\over\de p_\r}+{1\over2p_\r}\right),\qquad \hJ=i{\de\over\de p_\h},
\ee
where the scalar product
\be
<\q\ver\f>\ =\int_{-\inf}^{+\inf}{p_\r dp_\r dp_\h\over1+\b^2 p_\r^2}\ \q^*(p)\f(p)
\ee
is understood.
The completeness relations for momentum eigenstates are therefore
\be
\int_{-\inf}^{+\inf}{p_\r dp_\r\over1+\b^2 p_\r^2}\int_0^{2\p}dp_\h\ver p_\r,p_\h><p_\r,p_\h\ver=1
\ee

The eigenvalue equations for the position operators read\footnote{Also in this case the eigenfunction
are not physical because their energy diverges.}
\be
\hr\ver\r>\ =\r\ver\r>,\qquad\ver\r>\ \propto{\ex^{-i\r{\arctan\b p_\r\over\b}}\over\sqrt{p_\r}},
\ee
and
\be
\hJ\ver J>\ =J\ver J>,\qquad\ver J>\ \propto\ex^{-iJp_\h},
\ee
with integer $J$.

Defining a basis $\ver\r,J>\ =\ver\r>|\,J>$, one has
\be
<p_\r,p_\h\ver\r,J>\ ={\ex^{-i(\r{\arctan\b p_\r\over\b}+Jp_\h)}\over\sqrt{p_\r}}.
\ee
with
\be
\int_0^\inf d\r\int_{-\inf}^\inf dJ\ver\r,J><\r,J\ver=1.
\ee
Hence,
\be\lb{2dpq}
<\r,J\ver\ex^{-i\e H}\ver p_\r,p_\h>\ ={\ex^{i(\r{\arctan\b p_\r\over\b}+Jp_\h-\e H)}\over\sqrt{p_\r}}.
\ee

From (\ref{2dpq}), proceeding as in the one-dimensional case one obtains in the limit $\e\to0$ the
formula (\ref{pi}) with action (\ref{2dac}), and measure adapted to two dimensions,
\be\lb{mes2}
\cD p=\lim_{N\to\inf}\prod_{k=1}^{N-1}\int{d^2p^{(k)}\over1+\b^2p^{(k)\,2}},\qquad\cD q=\lim_{N\to\inf}\prod_{k=1}^{N}
\int d^2q^{(k)},
\ee
where $d^2p^{(k)}=p_\r^{(k)}dp_\r^{(k)}dp_\h^{(k)}$ and $d^2q^{(k)}=d\r^{(k)}dJ^{(k)}/p_\r^{(k)}$.

Moreover, in terms of the previous coordinates, the classical Hamiltonian $H={p_\r^2\over2}+V(r^2)$ takes the
form $H={p_\r^2\over2}+V\left(\r^2+{J^2\over p_\r^2}\right)$.
However, it is known that in order to obtain the correct result from the path integral, that takes
account of the hermitian nature of the operator $\hat\r^2$, an additional term $-{1/2 p_\r^2}$ must be added
to the classical two-dimensional action \cite{fu}. Hence, the correct effective potential will be
$V=V\left(\r^2+{J^2-1/2\over p_\r^2}\right)$.

\section{Faddeev-Jackiw formalism}
It is remarkable that the previous
 results can be obtained in an easier way using the first-order formalism introduced by Faddeev and
Jackiw. In fact, it is shown in \cite{FJ,jac}, using a Darboux transformation from the original variables
$\x_i$ to new canonical variables,  that the path integral can be written as
\be
A=\int\cD\x_i\,|\det \O_{ij}|^{-1/2}\ \ex^{iI},
\ee
where the determinant arises from the Jacobian of the transformation and $I$ is given by (\ref{fac}).

In our case, in any dimension,
\be
|\det \O_{ij}|=(1+\b^2p_i^2)^2,
\ee
from which the measures (\ref{mes1}) and (\ref{mes2}) follow.

Notice that this method can be easily employed to study different noncommutative models, including for example
the Moyal plane, investigated in \cite{ac} by different means.

\section{Two-dimensional examples}

In this section, we give a few elementary examples of application of the formalism to two-dimensional models.

In the case of a free particle, the integration over the angular variables $p_\h$ and $J$ simply yields a
delta function $\d(p_\h^{(i)}-p_\h^{(f)})$, and one is left with an integral over
the radial coordinates,
\be
\int\cD\r\int \cD p_\r\exp\left[i\int\left({\r\dot p_\r\over1+\b^2p_\r^2}+{p_\r^2\over2}\right)dt\right].
\ee
where
\be
\cD p_\r=\lim_{N\to\inf}\prod_{k=1}^{N-1}\int{d p_\r^{(k)}\over1+\b^2p_\r^{(k)\,2}}.
\ee
Performing a change of variables $P_\r=\b^\mo\arctan\b p_\r$, one gets
\be
\int\cD\r\int\cD P_\r\exp\left[i\int\left(\r\dot P_\r+{\tan^2 P_\r\over2}\right)dt\right],
\ee
with
\be
\cD P_\r=\lim_{N\to\inf}\prod_{k=1}^{N-1}\int d P_\r^{(k)}.
\ee
The integration on $\r$ gives in turn gives a delta function $\d(P_\r^{(i)}-P_\r^{(f)})$ and one is left with
\be
\int\cD P_\r\exp\left[{i\over2}\int\tan^2 P_\r\ dt\right].
\ee

In the harmonic oscillator case, the classical potential is $V=\o^2r^2\to\o^2\left(\r^2+{J^2-1/4\over p_\r^2}\right)$.
 It is now convenient to integrate first in $p_\h$, getting the conservation
of the angular momentum, $\d(J^{(i)}-J^{(f)})$. The integral reduces then to a sum on different $J$ sectors:
\be
\int\cD\r\int {\cD p_\r\over1+\b^2p_\r^2}\exp\left[i\int\left({\r\dot p_\r\over1+\b^2p_\r^2}+{p_\r^2\over2}+\o^2\r^2+
\o^2\,{J^2-1/4\over p_\r^2}\right)dt\right].
\ee
Again defining a new variable $P_\r$ as before, one gets
\be
\int\cD\r\int\cD P_\r\exp\left[i\int\left(\r\dot P_\r+\o^2\r^2+{\tan^2P_\r\over2}+\o^2\,{J^2-1/4\over\tan^2P_\r}\right)
dt\right],
\ee
and the gaussian integration over $\r$ yields
\be
\int\cD P_\r\exp\left[-i\int\left({\dot P_\r^2\over4\o^2}-{\tan^2P_\r\over2}-\o^2\,{J^2-1/4\over\tan^2P_\r}\right)dt
\right].
\ee
This path integral can be evaluated at least perturbatively by standard methods, and is similar to that obtained in the
one-dimensional case in \cite{val}.

\section*{Acknowledgement}

SM wishes to thank Petr Jizba for an interesting discussion.

\end{document}